\documentclass[preprint2]{aastex}

\shorttitle{Globular clusters as modified gravity probes}
\shortauthors{X. Hernandez and M. A. Jim\'enez}

\begin{document}


\title{Isothermal distributions in MONDian gravity as a simple unifying explanation for the ubiquitous
$\rho \propto r^{-3}$ density profiles in tenuous stellar halos}


\author{X. Hernandez, M. A. Jim\'enez and C. Allen}
\affil{Instituto de Astronom\'{\i}a, Universidad Nacional Aut\'{o}noma de M\'{e}xico,
Apartado Postal 70--264 C.P. 04510 M\'exico D.F. M\'exico}


\altaffiltext{1}{email:xavier@astro.unam.mx}


\begin{abstract}
That the stellar halo of the Milky Way has a density profile which to first approximation satisfies $\rho \propto r^{-3}$ 
has been known for a long time. More recently, it has become clear that M31 also has such an extended stellar halo, which 
approximately follows the same radial scaling. Studies of distant galaxies have revealed the same phenomenology. Also, 
we now know that the density profiles of the globular cluster systems of our Galaxy and Andromeda to first approximation 
follow $\rho \propto r^{-3}$, $\Sigma \propto R^{-2}$ in projection. Recently, diffuse populations of stars have been detected 
spherically surrounding a number of Galactic globular clusters, extending much beyond the Newtonian tidal radii, often without 
showing any evidence of tidal features. Within the standard Newtonian and GR scenario, numerous and diverse particular explanations 
have been suggested, individually tailored to each of the different classes of systems described above. Here we show that in a 
MONDian gravity scenario, any isothermal tenuous halo of tracer particles forming a small perturbation surrounding a spherically 
symmetric mass distribution, will have an equilibrium configuration which to first approximation satisfies a $\rho \propto r^{-3}$ scaling.
\end{abstract}



\section{Introduction}

Progress in the debate between the hypothetical physical reality of astrophysical dark matter and the option of modifying 
gravity at the low acceleration regime, will hinge upon the exploration of as many independent lines of enquiry as
possible. Whereas the rotation curves of galaxies can be adequately reproduced from either point of view (Milgrom 1983, Milgrom 1994, 
Sanders \& McGaugh 2002, Swaters et al. 2010), a variety of recent studies have shown results in tension with the standard 
scale-invariant gravity plus dark matter paradigm, and being more in line with generic MONDian gravity approaches. 

We shall use the term MONDian to refer to modified gravity theories in which, in the low velocity limit, the force per unit mass 
between a test particle and a spherical mass distribution will shift from the Newtonian expression of $G M/r^{2}$ to an 
$(G M a_{0})^{1/2}/r$ behaviour for acceleration scales below Milgrom's $a_{0}$, independently of the fundamental underlying theory 
of gravity which might lead to such a behaviour e.g. Bekenstein (2004), Moffat \& Toth (2008), Zhao \& Famaey (2010), Bernal et al. 
(2011), Mendoza et al. (2011), Capozziello \& De Laurentis (2011). The most salient features of such schemes are equilibrium velocities 
which become independent of distances at a value $\approx(G M a_{0})^{1/4}$.

In Lee \& Komatsu (2010) and Thompson \& Nagamine (2012) it has been shown that the infall velocity of the two components of the Bullet 
cluster is incompatible with expectations of full $\Lambda$CDM predictions, and is actually a challenge to the standard gravity
theory, as it surpasses the escape velocity of the combined system. Recently, Kroupa (2012) has shown that tidal dwarf galaxies,
which under the standard scenario are transient tidal clumps of galactic material having out of equilibrium dynamics, actually
fall on the same Tully-Fisher relation as all dwarf galaxies. This appears as an uncanny coincidence from the standard gravity
perspective, where the dynamics of normal dwarfs are thought to be determined by their dominant dark matter halos. The result is expected
under MONDian gravity schemes, where once the baryonic mass of the system is fixed, the dynamics will uniquely follow, as it is indeed
observed.

Along the same lines, in Scarpa et al. (2011), Hernandez \& Jim\'enez (2012) and Hernandez et al. (2013) it has been shown that the velocity
dispersion profiles of a number of Galactic globular clusters stop falling radially along Keplerian expectations and settle to finite
asymptotic values on crossing the $a<a_{0}$ threshold. The standard gravity explanation for these profiles, that it is the tides of the 
Milky Way that dynamically heat the outskirts of the clusters observed e.g. Lane et al. (2010), appears suspect, as the Newtonian tidal 
radii can be shown to exceed the points where the velocity dispersion profiles flatten by large factors, and because the total globular 
cluster masses and asymptotic velocity dispersion values follow the galactic Tully-Fisher relation, as expected under MONDian gravity 
schemes. Finally, in Hernandez et al. (2012) we showed that the relative velocities of extremely wide binaries do not follow the 
expectations of full galactic dynamical Newtonian simulations, but diverge from the Keplerian decline with radius to settle also at 
finite relative velocities on crossing the $a<a_{0}$ threshold of MONDian gravity proposals.

In this paper we show that the density profile of an isothermal population of trace particles surrounding a spherical mass distribution
in MONDian gravity will naturally follow an approximately $\rho \propto r^{-3}$ profile. Under standard gravity approaches, 
the ubiquitous nature of $\rho \propto r^{-3}$ profiles has to be addressed through a variety of highly specific explanations, 
individually tailored to each of the diverse classes of systems where these profiles have been observed. Examples of the above are 
mergers and tidal dissolution of accreted sub-structure for the tenuous stellar halos surrounding our Galaxy, M31 and also the recently 
detected ones around external galaxies (e.g. Bullock et al. 2001, Abadi 2006), and the compression of tidal tails at apocentre or disk 
shock heating for the "extra tidal" features surrounding Galactic globular clusters, e.g. Da Costa (2012).

The above situation contrasts with the appearance of a direct equilibrium solution for tracer populations having the simplest 
isotropic and isothermal Maxwellian distribution function, which to first order yields $\rho \propto r^{-3}$ profiles under 
MONDian gravity schemes, which we show here. We suggest this is one further piece of evidence pointing in the direction of the 
necessity of a modified gravity regime at low acceleration scales.

In Section (2) we derive the first order predictions for equilibrium tenuous stellar halos under a MONDian gravity scheme, and also
under Newtonian gravity within dark matter halos, and under Newtonian gravity in the absence of dark matter halos. In section (3) we
compare the theoretical estimates with the observed profiles for a variety of systems, finding the MONDian prediction a good fit to the
observed situation, over a wide variety of scales and classes of astronomical systems. Section (4) presents a final discussion of
our results.

\section{First Order Dynamical Expectations}

We shall model the physical situation which applies generically for a MONDian scenario, where in the $a<<a_{0}$ limit,  
the force between a test particle and a spherically symmetric mass distribution becomes $-(G M(r) a_{0})^{1/2}/r$, when
one introduces no modifications to Newton's second law e.g. Hernandez et al. (2010), Mendoza et al. (2011).

Assuming spherical symmetry, and taking the derivative of the kinematic pressure, the equation of hydrostatic equilibrium for a 
polytropic equation of state $P=K\rho^{\gamma}$ is:
  
\begin{equation}
\frac{d(K\rho^{\gamma})}{dr}= -\rho\nabla \phi.
\end{equation}
In going to isothermal conditions, $\gamma=1$ and $K=\sigma^2$. e.g. Binney \& Tremaine (1987), and we get: 

\begin{equation}
\frac{\sigma^{2}}{\rho} \frac{d\rho}{dr}= -\nabla \phi.
\end{equation}

writing  $\rho=(4\pi r^2)^{-1}\frac{dM(r)}{dr}$ the above equation can be written as:

\begin{equation}
\sigma^{2} \left[\left(\frac{dM(r)}{dr}\right)^{-1}\frac{d^{2}M(r)}{dr^{2}}-\frac{2}{r} \right]\nonumber =-\nabla \phi(r)
\label{eqhidro}
\end{equation}

\noindent 
where $\sigma$ is the isotropic Maxwellian velocity dispersion for the population of stars. The above treatment is common, and can 
be found in e.g. Hernandez et al. (2010), where we used it in the modeling of dSph galaxies, systems characterised by flat 
velocity dispersion profiles, obtaining mass models consistent with observed velocity dispersion, half mass radii and 
total masses, in the absence of dark matter. Other recent examples of similar treatments can found in e.g. Drukier et al. (2007),
Sollima \& Nipoti (2010) and Hernandez \& Jim\'enez (2012), all modelling stellar populations using isotropic Maxwellian distribution 
functions.

As an illustrative example we can take $\nabla \phi(r) = G M(r)/r^{2}$ for the right hand side of
equation (3), the Newtonian expression appearing for $a>> a_{0}$. Looking for a power law solution 
for $M(r)=M_{0}(r/r_{0})^{m}$, we get:

\begin{equation}
\sigma^{2} \left[ \frac{m-3}{r}\right]
=-\frac{ G M_{0}}{r^{2}} \left( \frac{r}{r_{0}} \right)^{m}, 
\end{equation}

\noindent and hence $m=1$, the standard isothermal halo, $M(r)=2 \sigma^{2} r/G$, having a constant centrifugal equilibrium 
velocity $v^{2}=2 \sigma^{2}$ and infinite extent. In going to the MONDian limit of $a<<a_{0}$, $\nabla \phi (r) = (G M(r) a_{0})^{1/2}/r$, 
equation (3) yields:

\begin{equation}
\sigma^{2} \left[  \frac{m-3}{r}     \right]
=-\frac{[G M_{0} a_{0}]^{1/2}}{r} \left( \frac{r}{r_{0}} \right)^{m/2}.
\end{equation}

\noindent In this limit $m=0$, we obtain
$M(r)=M_{0}$ and $v^{2}= 3 \sigma^{2}=(G M_{0} a_{0})^{1/2}$, the expected Tully-Fisher scaling of the circular
equilibrium velocity with the fourth root of the mass, with rotation velocities which remain flat even after the mass
distribution has converged, thus, rigorously isothermal halos are naturally limited in extent, as already shown by Milgrom (1984). 
It is interesting that in this limit the scaling between the circular rotation velocity and the velocity dispersion is only slightly 
modified as compared to the Newtonian case, with the proportionality constant changing from 2 to 3, for the squares of the velocities.
Notice also that a fuller asymptotic analysis (Milgrom 1984) not imposing a power law solution, shows the factor of 3 obtained
above will in general lie in the range 3-4.5.

We can now look for the behaviour of a tenuous stellar halo in the MONDian regime, and hence at large distances, around a mass distribution 
which has essentially converged, by looking at eq.(2) and writing the R.H.S. as $-(G M_{tot} a_{0})^{1/2}/r$, where $M_{tot}$ is the total 
mass of the galactic or stellar system. The $\rho$ in the L.H.S. of this same equation now refers to the density distribution of essentially
test particles making up a trace population, e.g. a stellar galactic halo, the globular cluster distribution around a large galaxy, or the 
faint halos of ``extra-tidal'' stars surrounding Galactic globular clusters. Using also the result of eq.(5) of 
$3 \sigma^{2}=(G M_{0} a_{0})^{1/2}$, eq.(2) yields:

\begin{equation}
\frac{d \rho}{d r} =-3 \frac{\rho}{r},
\end{equation}

\noindent which can then be integrated directly to yield:

\begin{equation}
\rho(r)=\rho_{0} (r_{0}/r)^{3}. 
\end{equation}

In deriving eq.(7) we have introduced the assumptions of a tracer population and the results of having forced a power
law solution in eq.(5), which significantly simplify the calculations with respect to a full numerical solution (e.g. Milgrom 1984 
or Hernandez \& Jim\'enez 2012), or even with respect to the asymptotic analysis of Milgrom (1984). The above assumptions will 
certainly never be strictly valid in a real astrophysical system, still, provided they are approximately valid, the solution of eq.(7) 
will represent a first order description. Our simplified approach however, allows a transparent handling of the physics, and permits
a clear understanding for the generic appearance of a $\rho(r) \propto r^{-3}$ region, a feature already noticed in the numerical solutions
presented in Milgrom (1984), and apparent in many astrophysical systems, as discussed in the following section.    
The volumetric distribution of eq.(7) can be projected analytically along one direction to yield the projected surface density profile 

\begin{equation}
\Sigma(R)=\frac{\pi \rho_{0} r_{0}^{3}}{2 R^{2}}.
\end{equation}

Of course, $\rho\propto r^{-3}$ is an approximation which will only be valid over a limited radial range. In fact, mass profiles 
for isothermal solutions converge to finite total masses and radii, as shown in e.g. Milgrom (1984), Hernandez et al. (2010).
In closer detail, the density profiles will steepen beyond $r^{-3}$ as one moves further out as the total mass converges, e.g. 
as seen in the broken power law fits for the Galactic stellar halo reported by Sesar et al. (2011).

Under the Newtonian expression of $\nabla \phi= GM(r)/r^{2}$, the equivalent development for a trace population in the halo of a galaxy
having the same rotation curve as the one leading to eq.(6), $M(r)=2 \sigma^{2} r/G$, where this time $M(r)$ refers mostly to the 
hypothetical dark matter component, yields: 

\begin{equation}
\rho(r) = \rho_{0} (r_{0}/r)^{2}
\end{equation}

\noindent as the expression corresponding to eq.(7). The case of a trace population around an essentially converged total mass in 
Newtonian dynamics, e.g. the tenuous stellar halos surrounding the Galactic globular clusters, $\nabla \phi= GM_{tot}/r^{2}$ yields:

\begin{equation}
\rho(r)=\rho_{0} e^{(G M_{tot}/\sigma^{2} r)},
\end{equation}

\noindent a density distribution which tends to a constant at large radii. Thus, we see that equilibrium configurations of isothermal 
tracer populations in the MONDian regime will approximately follow $\rho(r) \propto r^{-3}$ density profiles, while under Newtonian 
gravity the same populations within the corresponding dark matter halos will show much shallower $\rho(r) \propto r^{-2}$ profiles, 
which in the absence of dark matter halos e.g. tenuous stellar halos about globular clusters, will have density profiles as given by eq.(10).

\section{Observational comparisons}

In this section we review the observational situation of tenuous tracer population halos, which now spans a very wide range of systems 
and astrophysical scales. We begin with a number of recent determinations of the density structure of the stellar halo of our Galaxy. 
Morrison et al. (2000) implement a careful disk/halo star separation criteria, and obtain $\rho(r) \propto r^{-3}$ for the stellar halo 
of the Milky Way. Juric et al. (2008) report a single power law fit $\rho(r) \propto r^{-2.8 \pm 0.3}$, while Bell et al. (2008) find halo 
profiles having more structure than simple power laws to yield better fits, but still, $\rho(r) \propto r^{-3}$ for the preferred single 
power law model. Finally, Sesar et al. (2011) obtain a broken power law as the most accurate description, but again, a best fit single 
power law of $\rho(r) \propto r^{-2.9}$. There is clearly a broad radial range over which the best fit single power law model 
yields a slope as expected from eq. (7).

Recent detailed studies of the stellar halo of Andromeda using a variety of techniques and data from the largest modern facilities 
have reached a consensus for a $\rho \propto r^{-3}$ structure. Ibata et al. (2007) obtain $\Sigma(R) \propto R^{-1.91 \pm 0.12}$ with 
data from the CFHT, Tanaka et al. (2010) using the Suprime-Cam instrument on the Subaru telescope found 
$\Sigma(R) \propto R^{-2.17 \pm 0.15}$, and lastly Gilbert et al. (2012) measure $\Sigma(R) \propto R^{-2.2 \pm 0.2}$ out to very large 
distances, 175 kpc, coming to $\Sigma(R) \propto R^{-2.0 \pm 0.5}$ for $20kpc< R< 90kpc$ once the kiematical substructure is removed. 
It appears that the stellar halo of M31 is a classic example of the tenuous populations described by eq. (7).

Regarding more external galaxies, tenuous extended stellar halos have been detected over the past few years surrounding numerous systems.
Recent detections include Cockcroft et al. (2012) who find evidence for an extended stellar halo about M33 and Bakos \& Trujillo (2012) 
who report finding such structures about a sample of 7 late-type spirals from the SDSS, in all cases with total masses amounting to only a 
few percent of the total baryonic mass of the host galaxies. Although the above observations can not yet yield secure projected density 
profiles, Jablonka et al. (2010) report a tenuous stellar halo about NGC 3957 with a projected scaling $\Sigma(R) \propto R^{-2.76 \pm 0.43}$,
while Baker et al. (2009) find a faint stellar halo about M81 with a projected scaling $\Sigma(R) \propto R^{-2.0 \pm 0.2}$, and Bailin et al. 
(2011) observe a stellar halo about NGC 253 with $\Sigma(R) \propto R^{-2.8 \pm 0.6}$. In going to larger samples, Zibetti et al. (2004) 
showed through the stacking of images from 1047 edge-on spiral galaxies from the SDSS, that these very generally present extended tenuous 
stellar halos with volumetric radial density profiles well described by $\rho(r) \propto r^{-3}$.

In going to the spatial distribution of a different tracer population, this time the globular cluster systems of galaxies, it has
been well know for many years that the density profile of the GC system of the Milky Way very accurately follows a $\rho(r)\propto r^{-3}$
profile, e.g. Surdin (1994), Racine \& Harris (1998). Looking in more detail, more recently Bica et al. (2006) find a  
$\rho(r)\propto r^{-n}$ profile with $3.2<n<3.9$ for all Galactic globular clusters, while the metal rich population follows also a
power law, this time with $\rho(r)\propto r^{-3.2 \pm 0.2}$ for large radii, and $\rho(r)\propto r^{-3.2 \pm 0.9}$ if one includes the effects
of oblateness in the distribution. The globular cluster system of M31 has a projected power law density profile also in agreement
with the expectations of eq.(7), e.g. Racine (1991) determined $\Sigma(R) \propto R^{-2}$. More recently and in more detail, Huxor et al. 
(2011) obtained a best fit profile composed of three distinct power laws, which however, if modelled as a single power law for $r>1$kpc, 
can be approximated by the same $\Sigma(R) \propto R^{-2}$ law found earlier by Racine (1991).

Going to more external galaxies, Perelmuter \& Racine (1995) found a best fit $\Sigma(R) \propto R^{-2}$ scaling for the globular cluster 
system of M81. Harris et al. (1984) found the outer projected radial distribution of globular clusters in NGC 4594, the Sombrero galaxy, 
to be well described by a $\Sigma(R) \propto R^{-2}$ profile. Harris and van den Bergh (1981) found also $\rho(r) \propto r^{-3}$ scalings 
for the globular cluster systems around 7 elliptical galaxies. More recent studies have found a spread in the power law slopes of projected 
density profiles for globular cluster systems, but taken as a whole, ``typical projected power-law indices range from −2 to −2.5 for some 
low-luminosity Es to −1.5 or a bit lower for the most massive giant ellipticals.'' to quote from the review from Brodie \& Strader (2006).

Since the studies of Grillmair et al. (1995) and Leon et al. (2000), a number of tenuous stellar halos associated to 
Galactic globular clusters have been detected. The problem of determining structural parameters is harder than in the cases of the stellar 
halos surrounding galaxies, as the overall numbers of stars are much lower, and the problem of contamination by foreground and background 
sources, as well as by obscuration, is significant. More modern studies have found a large range of power law slopes in the outskirts of 
globular clusters, McLaughlin \& van der Marel (2005) find projected indexes going from -2 to -6, but do report that a population of the 
most massive clusters shows indexes close to $\Sigma(R) \propto R^{-2}$, and conclude that the extended halos enveloping the clusters they 
study are suggestive of a generic equilibrium feature, rather than being transient structures. Jordi \& Grebel (2010) report projected 
power law indexes for tenuous stellar halos surrounding 17 Galactic globular clusters; their most reliable results span values from -1 to -4. 
These authors also note features which are problematic for a standard gravity interpretation, in the cases of e.g. NGC 7089 and Pal 1, whose 
stellar halos are clearly spherical with no sign of any tidal features, in spite of extending in both cases much beyond their Newtonian 
Jacobi radii. Carballo-Bello et al. (2012) perform a similar study (also including careful CMD modelling to limit contamination) for the 
extra-tidal halos of 19 Galactic globular cluster. Again, reported projected power law indexes span a broad range from close to -2 to about 
-4, excluding clusters showing clear tidal features. Note that for two of the three clusters which overlap with the Jordi \& Grebel (2010) 
sample, NGC 4147 and NGC 5272, Carballo-Bello et al. (2012) report power law indexes of $-2.8^{+0.07}_{-0.05}$ and $-3.18^{+0.08}_{-0.05} $, 
while Jordi \& Grebel (2010) assign to these same clusters values of $-1.48 \pm 0.24$, and of $-0.94 \pm 0.38$ respectively, for a comparable 
radial range. This simply illustrates that the observational situation is far from converging to definitive answers regarding these systems.

Further, in the case of globular clusters the problem is intrinsically less clear than for the galactic stellar halos, as internal 
dynamical evolutionary effects might play a part, as well as the gravitational perturbations due to the crossing of the Galactic disk. 
It is however clear that Galactic globular clusters often, if not always, are surrounded by tenuous stellar halos, which from the 
Newtonian point of view, must be regarded as ``extra tidal'' structures. The smooth and round appearances often observed, with all 
absence of tidal tails, hence become a problem. This problem does not appear under MONDian gravity, where satellite systems are generically 
expected to be much more robust to tides, e.g. Hernandez \& Jim\'enez (2012). Interestingly, Mackey et al. (2010) find a 
$\Sigma(R) \propto R^{-n}$ power law structure for the tenuous stellar halo surrounding an extremely isolated globular cluster in M31, 
which begins as $n \approx 2.5$, and then breaks further outwards to $n \approx 3.5$. Also, note that most of the $\Sigma(R) \propto R^{-n}$ 
indexes in the two recent Jordi \& Grebel (2010) and Carballo-Bello et al. (2012) studies cluster about $n=3$. The presence of as yet 
undiscovered tidal features among these clusters would tend to artificially steepen their profiles.  Note also that already Grillmair et al. 
(1995) cautioned that the difficulties of background subtraction and obscuration corrections will lead to systematics which tend to yield 
overestimates in $n$. On the other hand, the large indexes sometimes reported for globular clusters could also be detections of the 
steepening in the profile expected under MONDian gravity models on approaching the final radius, already mentioned in the discussion 
following eq.(8).

It thus appears clear that extended tracer population halos, stars or globular clusters, having a small fraction of the light of their 
host systems, galaxies or globular clusters, are a common feature. Also, such halos are generally never far from the predictions of 
eq.(7) for equilibrium configurations of isothermal tracer populations in the MONDian regime, to first approximation 
$\rho(r) \propto r^{-3}$ or $\Sigma(R) \propto R^{-2}$. Within the standard gravity interpretation, explanations to the power law 
density profiles of galactic stellar halos have been proposed in terms of the accretion and tidal dissolution of substructure falling 
into the main galaxy e.g. Bullock et al. (2001), Bullock \& Johnston (2005), and Abadi et al. (2006). However, the overall smoothness 
and uniformity in stellar properties of these systems has been pointed out as problematic for the standard explanation, which naturally 
implies a degree of randomness in the accreted material, e.g. Ibata et al. (2007), and Bell et al. (2008), who also find from simulations 
stellar halos not matching observations in terms of the substructure details. Ibata et al. (2007) also show that standard simulations 
sometimes yield exponents of $\rho(r) \propto r^{-n}$ inconsistent with observations, with $n=4$ or even $n=5$. Also, a further explanation 
must be sought for the observed profiles of the globular cluster systems surrounding galaxies, and yet another for the  remarkably 
smooth ``extra-tidal'' stellar halos surrounding many of the globular clusters of the Milky Way.

\section{Discussion}

The relevance of the $\rho(r) \propto r^{-3}$ solution presented here to the observations listed above depends crucially on the validity 
of three assumptions regarding the tracer population in question, and which enter into the derivation of eq.(7): (i) that it lies within 
the $a<a_{0}$ region over which the modified gravity regime is thought to apply; (ii) that its velocity dispersion does not depend on 
radius, i.e. that it is isothermal; and lastly, (iii) that there is no orbital anisotropy present in its velocity dispersion, i.e., 
that it is isotropic. The validity of the first assumption is easy to verify; the radial ranges over which galactic stellar halos and 
globular cluster populations are observed to comply with the $\rho(r) \propto r^{-3}$ profiles are within the radial ranges where flat 
rotation curves are seen, and hence, from the accurate rotation curve modelling which MOND affords (e.g. Swaters et al. 2010), also 
within the $a<a_{0}$ region. In the case of the tenuous stellar halos surrounding Galactic globular clusters, these appear at radial 
distances comparable to, but mostly larger than, the regions where the $a<a_{0}$ threshold is crossed and the velocity dispersion 
profiles flatten, in the cases where this last have been measured, e.g. Scarpa et al. (2011), Hernandez et al. (2013). 

Regarding the second assumption, in all of the cases listed in the previous section, wherever a radial profile has been measured for 
the velocity dispersion of the tracer populations in question, these have been shown to be consistent with a constant isothermal solution. 
Examples of this last point are Battaglia et al. (2006) who from a sample of 240 halo objects obtain a velocity dispersion profile for 
the halo stars in the Milky Way consistent, within errors, with a constant value from 15 kpc to about 70 kpc, and with an inferred 
anisotropy consistent with an isotropic distribution. Brown et al. (2010) do find a falling trend for the velocity dispersion profile 
of stars in the MW halo, but only a very mild radial drop, while more recently Samurovic \& Lalovic (2011) obtain a velocity dispersion 
profile for a large sample of 2557 BHB stars in the MW from Xue et al. (2008) which is consistent with a constant value out to 70 kpc. 
Finally, Kafle et al. (2012) using 4664 blue horizontal branch stars from Xue et al (2011) in the MW halo observe a radial velocity 
dispersion which is indistinguishable from flat outwards of about 15 kpc, out to their last measured point at close to 60 kpc. Also, 
in all cases where the velocity dispersion profiles of stars in Galactic globular clusters have been measured out to large radii, 
these can be seen to be consistent with constant $\sigma$ values, e.g. Scarpa et al. (2011), Hernandez et al. (2013). An interesting 
feature of the two most recent references measuring the velocity dispersion profile of the stellar halo of our Galaxy listed above, 
is that the level for the constant 1D velocity dispersion found is of between 100 and 110 $km/s$, which would bring it in accordance 
with the expectations of the $3 \sigma^{2} =v^{2}$ condition we derive in section (2), since $\sqrt{3} \times 110=190$, the observed 
asymptotic rotation velocity of the MW.

The third assumption is much harder to test empirically, as no reliable measurements of orbital anisotropy exist for any of the 
astronomical systems treated here. Orbital isotropy is however a natural first order approximation commonly used in the modelling 
of self-gravitating systems, e.g. Binney \& Tremaine (1987), or Drukier et al. (2007), Sollima \& Nipoti (2010) and Hernandez \& 
Jim\'enez (2012) in the modelling of globular clusters under either Newtonian or MONDian approaches. Although an idealisation, it 
provides a convenient reference solution for a variety of dynamical studies of self-gravitating systems; for instance, studies of 
dynamical friction due to both the hypothetical dark matter and stellar components of dSph galaxies, routinely assume isotropic distribution
functions for the stars in question, even though this assumption is understood as only a first order approximation (e.g. Sanchez-Salcedo 
et al. 2006, Goerdt et al. 2006 and Cole et al. 2012, to cite a few recent examples). In the absence of any evidence suggesting orbital 
anisotropy for the systems and radial ranges treated here, e.g. any observed dominant flattening in the light distribution, we chose not 
to introduce any at this initial point. Further, in the particular case of MOND gravity, it was already shown in Milgrom (1984), that 
the introduction of a slight degree of anisotropy, which could in principle be present, modifies only slightly the resulting density 
profiles of self-gravitating systems. 

It is of course true that under Newtonian gravity, a $\rho(r) \propto r^{-3}$ profile for a tracer population can also be found, but 
only if one allows for more complex $\sigma(r)$ and radially varying anisotropy parameters. Since the observed density profiles for 
tenuous halos match the simplest isothermal (as observed in all cases where this function has been measured) and isotropic distribution 
functions under MONDian gravity, this solution is to be preferred to the fitting of contrived, ad hoc, radial variations in the velocity 
dispersion and anisotropy parameters of the tracer populations in question under Newtonian gravity, especially as none such variations 
have been detected. 

Clearly, for any astrophysical system where the halo population ceases to be a small perturbation on the
total mass, or where the velocity dispersion profile is seen to deviate significantly from the isothermal condition assumed here,
our solution will not be relevant. In spite of the approximate nature of the solution (due to the tracer population
assumption, the strict isothermal assumption and the forcing of a power law solution), the approximately isothermal profile of 
various systems where this has been observed, the very low mass contribution of the tracer populations treated, and the good match
to a $r^{-3}$ density profile which a wide variety of systems present, give us confidence in that some of the physics has been 
captured by the modelling.

Finally, it is interesting that some of the systems mentioned in the previous section are not in the deep MOND regime, therefore,
from the point of view strictly of MOND as such, no significant modifications to gravity should be apparent. We note that
the external field effect of MOND will be substantially modified for different modified gravity theories, of the various types 
listed in the introduction. Indeed, MOND variants have been discussed where the external field effect is substantially reduced, 
or even practically disappears, e.g. Milgrom (2011). We note also that our previous results of Hernandez et al. (2012) looking 
at the observed relative velocities of wide binaries in the solar neighbourhood, or of Hernandez et al. (2013) finding MONDian 
phenomenology in the observed outer dynamics of Galactic globular clusters, both classes of systems not in the deep MOND regime, 
strongly suggest a modified gravity theory where no external field effect appears. 

To summarise, we have shown that under a MONDian gravity force law, the density profiles of isothermal tenuous tracer 
population halos with isotropic Maxwellian velocities surrounding spherical mass distributions will be well approximated 
by $\rho \propto r^{-3}$ scalings. We suggest that such equilibrium configurations provide a natural, and certainly general, 
explanation for the observed close to $\rho \propto r^{-3}$ behaviour of: the stellar halos surrounding the Milky Way, M31, 
and a variety of external galaxies, the density profiles of the globular cluster systems in our Galaxy and Andromeda, and 
the radial structure of the ``extra tidal'' stellar halos recently observed surrounding a number of Galactic globular clusters. 

\section{Acknowledgements}

The authors thank an anonymous referee for pointing out a number of relevant details which were not sufficiently 
clear in the original version. Xavier Hernandez acknowledges financial assistance from UNAM DGAPA grant IN103011. 
Alejandra Jim\'enez acknowledges financial support from a CONACYT scholarship.


\begin{thebibliography}{}


\bibitem[A(1)]{} Abadi M. G., Navarro J. F., Steinmetz M., 2006, MNRAS, 365, 747

\bibitem[A(2)]{} Bailin J., Bell E., Chappell S. N., Radburn-Smith D. J., De Jong R. S., 2011, ApJ, 736, 24

\bibitem[A(3)]{} Bakos J., Trujillo I., 2012, arXiv:1204.3082

\bibitem[A(4)]{} Battaglia G., 2006, MNRAS, 370, 1055

\bibitem[A(5)]{} Bell E. F., et al., 2008, ApJ, 680, 295

\bibitem[A(6)]{} Bernal T., Capozziello S., Hidalgo J. C., Mendoza S., 2011, Eur. Phys. J. C, 71, 1794

\bibitem[A(7)]{} Bekenstein J. D., 2004, Phys. Rev. D, 70, 083509

\bibitem[A(8)]{} Bica E., Bonatto C., Barbuy B., Ortolani S., 2006, A\&A, 450, 105

\bibitem[A(9)]{} Binney J., Tremaine S., 1987, Galactic Dynamics (Princeton University Press, Princeton, NJ)

\bibitem[A(10)]{} Brodie J. P., Strader J., 2006, ARA\&A, 44, 193

\bibitem[A(60b)]{} Brown W. R., Geller M. J., Kenyon S. J., Diaferio A., 2010, AJ, 139, 59

\bibitem[A(11)]{} Bullock J. S., Kravtsov A. V., Weinberg D. H., 2001, ApJ, 548, 33

\bibitem[A(12)]{} Bullock J. S., Johnston K. V., 2005, ApJ, 635, 931

\bibitem[A(13)]{} Carballo-Bello J. A., Gieles M., Sollima A., Koposov S., Martínez-Delgado D., Pe\~{n}arrubia J., 2012, MNRAS, 419, 14

\bibitem[A(14)]{} Capozziello S., De Laurentis M., 2011, Phys. Rep. 509, 167

\bibitem[A(15)]{} Cockcroft R., et al., 2012, MNRAS, in press

\bibitem[A(16)]{} Cole D. R., Dehnen W., Read J. I., Wilkinson M, I., 2012, MNRAS, 426, 601

\bibitem[A(17)]{} Da Costa G. S., 2012, ApJ, 751, 6

\bibitem[A(18)]{} Drukier G. A., Cohn H. N., Lugger P. M., Slavin S. D., Berrington R. C., Murphy B. W., 2007, AJ, 133, 1041

\bibitem[A(19)]{} Famaey B., McGaugh S., 2012, Living Reviews in Relativity, in press, arXiv:1112.3960
 
\bibitem[A(20)]{} Gilbert K. M., et al., 2012, ApJ, 760, 76 

\bibitem[A(21)]{} Goerdt T., Moore B., Read J. I., Stadel J., Zemp M., 2006, MNRAS, 368, 1073

\bibitem[A(22)]{} Grillmair C. J., Freeman K. C., Irwin M., Quinn P. J., 1995, AJ, 109, 2553

\bibitem[A(23)]{} Haghi H., Baumgardt H., Kroupa P., Grebel E. K., Hilker M., Jordi K., 2009, MNRAS, 395, 1549

\bibitem[A(24)]{} Haghi H., Baumgardt H., Kroupa P., 2011, A\&A, 527, A33

\bibitem[A(25)]{} Harris W. E., van den Bergh S., 1981, AJ, 86, 1627

\bibitem[A(26)]{} Harris W. E., Harris H. C., Harris G. L. H., 1984, AJ, 89, 216

\bibitem[A(27)]{} Hernandez X., Mendoza S., Suarez T., Bernal T., 2010, A\&A, 514, A101

\bibitem[A(28)]{} Hernandez X., Jim\'enez M. A., Allen C., 2012, EPJC, 72, 1884

\bibitem[A(29)]{} Hernandez X., Jim\'enez M. A., 2012, ApJ, 750, 9

\bibitem[A(30)]{} Hernandez X., Jim\'enez M. A., Allen C., 2013, MNRAS, 428, 3196

\bibitem[A(31)]{} Huxor A. P., et al., 2011, MNRAS, 414, 770

\bibitem[A(32)]{} Ibata R., Martin N. F., Irwin M., Chapman S., Ferguson A. M. N., Lewis G. F., McConnachie A. W., 2007, ApJ, 671, 1591

\bibitem[A(33)]{} Jablonka P., Tafelmeyer M., Courbin F., Ferguson A. M. N., 2010, A\&A, 513, A78

\bibitem[A(34)]{} Jordi K., Grebel E. K., 2010, A\&A, 522, A71

\bibitem[A(35)]{} Juric M., et al. 2008, ApJ, 673, 864 

\bibitem[A(35b)]{} Kafle P R., Sharma S., Lewis G. F., Bland-Hawthorn J., 2012, ApJ, 761, 98

\bibitem[A(36)]{} Kroupa P. et al., 2010, A\&A, 523, 32 

\bibitem[A(37)]{} Kroupa P., 2012, PASA, 29, 395
 
\bibitem[A(38)]{} K\"{u}pper A. H. W., Kroupa P., Baumgardt H., Heggie D. C., 2010, MNRAS, 407, 224

\bibitem[A(39)]{} Lane R. et al., 2010, MNRAS, 406, 2732

\bibitem[A(40)]{} Lee J., Komatsu E., 2010, ApJ, 718, 60

\bibitem[A(41)]{} Mackey A. D., et al., 2010, MNRAS, 401, 533 

\bibitem[A(42)]{} McLaughlin D. E., van der Marel, R. P.,2005, ApJS, 161, 304

\bibitem[A(43)]{} Mendoza S., Hernandez X., Hidalgo J. C., Bernal T., 2011, MNRAS, 411, 226 

\bibitem[A(44)]{} Milgrom M., 1983, ApJ, 270, 365

\bibitem[A(45)]{} Milgrom M., 1984, ApJ, 287, 571

\bibitem[A(46)]{} Milgrom M., 1994, ApJ, 429, 540

\bibitem[A(46a)]{} Milgrom M., 2011, arXiv:1111.1611

\bibitem[A(47)]{} Moffat J. W., Toth V. T., 2008, ApJ, 680, 1158

\bibitem[A(48)]{} Morrison H. L., Mateo M., Olszewski E. W., Harding P., Dohm-Palmer R. C., 
Freeman K. C., Norris J. E., Morita M., 2000, AJ, 119, 2254

\bibitem[A(49)]{} Perelmuter J. M., Racine R., 1995, AJ, 109, 1055

\bibitem[A(50)]{} Racine R., 1991, AJ, 101, 865

\bibitem[A(51)]{} Racine R., Harris W. E., 1989, AJ, 98, 1609

\bibitem[A(51b)]{} Samurovic S., Lalovic A., 2011, A\&A, 531, A82

\bibitem[A(52)]{} Sanchez-Salcedo F. J., Reyes-Iturbide J., Hernandez X., 2006, MNRAS, 370, 1829

\bibitem[A(53)]{} Sanders R. H., McGaugh S. S., 2002, ARA\&A, 40, 263

\bibitem[A(54)]{} Scarpa R., Marconi G., Carraro G., Falomo R., Villanova S., 2011, A\&A, 525, A148

\bibitem[A(55)]{} Sesar B., Juric M., Ivezic Z., 2011, ApJ, 731, 4

\bibitem[A(56)]{} Sollima A., Nipoti C., 2010, MNRAS, 401, 131

\bibitem[A(57)]{} Surdin V. G., 1994, Astronomy Letters, 20, 398

\bibitem[A(58)]{} Swaters R. A., Sanders R. H., McGaugh S. S., 2010, ApJ, 718, 380

\bibitem[A(59)]{} Tanaka M., Chiba M., Komiyama Y., Guhathakurta P., Kalirai J. S., Iye M., 2010, ApJ, 708, 1168

\bibitem[A(60)]{} Thompson R., Nagamine K., 2012, MNRAS, 419, 3560

\bibitem[A(61)]{} Zhao H., Famaey B., 2010, Phys. Rev. D, 81, 087304

\bibitem[A(62)]{} Zibetti S., White S. D. M., Brinkmann J., 2004, MNRAS, 347, 556

\bibitem[A(63)]{} Xue X. X., Rix H. W., Zhao G., et al., 2008, ApJ, 684, 1143

\bibitem[A(64)]{} Xue X. X., Rix H. W., Yanni B., et al., 2011, ApJ, 738, 79


\end{thebibliography}
\end{document}